\documentclass{article}

\usepackage{arxiv}

\usepackage[utf8]{inputenc} 
\usepackage[T1]{fontenc}    
\usepackage{hyperref}       
\usepackage{url}            
\usepackage{booktabs}       
\usepackage{amsfonts}       
\usepackage{nicefrac}       
\usepackage{microtype}      
\usepackage{graphicx}
\usepackage{natbib}
\usepackage{doi}
\usepackage{xcolor}
\usepackage[utf8]{inputenc}
\usepackage{algpseudocode}
\usepackage{color,soul}
\usepackage{subcaption}
\usepackage{adjustbox}

\usepackage{algorithm}

\usepackage{amsmath}

\usepackage{hypcap} 

\newtheorem{theorem}{Theorem}

\usepackage{makecell}

\usepackage{tabularray}

\title{NeuroAssist: Enhancing Cognitive-Computer Synergy with Adaptive AI and Advanced Neural Decoding for Efficient EEG Signal Classification}

\author{Eeshan G. Dandamudi\\
	neuralyx.ai\\
	\texttt{egdandamudi@gmail.com} \\
}

\renewcommand{\shorttitle}{\textit{arXiv} Template}

\hypersetup{
pdftitle={A template for the arxiv style},
pdfsubject={q-bio.NC, q-bio.QM},
pdfauthor={David S.~Hippocampus, Elias D.~Striatum},
pdfkeywords={First keyword, Second keyword, More},
}

\begin{document}

\maketitle

\begin{abstract}
Traditional methods of controlling prosthetics frequently encounter difficulties regarding flexibility and responsiveness, which can substantially impact people with varying cognitive and physical abilities. Advancements in computational neuroscience and machine learning (ML) have recently led to the development of highly advanced brain-computer interface (BCI) systems that may be customized to meet individual requirements. To address these issues, we propose NeuroAssist, a sophisticated method for analyzing EEG data that merges state-of-the-art BCI technology with adaptable artificial intelligence (AI) algorithms. NeuroAssist's hybrid neural network design efficiently overcomes the constraints of conventional EEG data processing. Our methodology combines a Natural Language Processing (NLP) BERT model to extract complex features from numerical EEG data and utilizes LSTM networks to handle temporal dynamics. In addition, we integrate spiking neural networks (SNNs) and deep Q-networks (DQN) to improve decision-making and flexibility. Our preprocessing method classifies motor imagery (MI) one-versus-the-rest using a common spatial pattern (CSP) while preserving EEG temporal characteristics. The hybrid architecture of NeuroAssist serves as the DQN's Q-network, enabling continuous feedback-based improvement and adaptability. This enables it to acquire optimal actions through trial and error. This experimental analysis has been conducted on the GigaScience and BCI-competition-IV-2a datasets, which have shown exceptional effectiveness in categorizing MI-EEG signals, obtaining an impressive classification accuracy of 99.17\%. NeuroAssist offers a crucial approach to current assistive technology by potentially enhancing the speed and versatility of BCI systems.

\end{abstract}


\section{Introduction}

Brain-computer interfaces (BCIs) use signals from the brain to control external devices, offering substantial potential for assistive technology, medical rehabilitation, and interactive gaming (\cite{camargo2021brain}. BCIs, although making great progress, are typically limited in their capacity to adapt to users' changing cognitive states and motor intents in real time, which hinders their overall usefulness. Conventional BCIs primarily depend on fixed algorithms that do not adapt to the changing patterns of human brain activity, resulting in less-than-ideal performance and user dissatisfaction (\cite{kawala2021summary}). The difficulty is further increased in applications that need exceptional accuracy and natural interaction, such as prosthetic control. Current approaches fail to adequately capture the complex patterns of motor imagery signals (\cite{orban2022review}). This situation emphasizes the need for creative methods to analyze and comprehend electroencephalography (EEG) data in real-time, guaranteeing that BCI adapts to user requirements and provides improved responsiveness and precision.

In considering the difficulties encountered by persons utilizing prostheses, it is crucial to acknowledge the significance of the problem. Based on statistical data, it is estimated that approximately 2 million individuals in the United States are amputees, with around 185,000 individuals utilizing prosthetic devices to replace lost limbs. Projections suggest that by the year 2050, the population of amputees in the United States will exceed 3.6 million, indicating a significant increase in the prevalence of limb loss and the demand for prosthetic solutions. A survey of 97 participants, comprising 60 men and 37 women, revealed insights into the satisfaction levels and preferences regarding prosthetic sockets. Among the 12 participants who provided feedback, 41.7\% expressed satisfaction, while 25\% reported feeling somewhat satisfied with their prosthetic sockets. Participants identified durability and comfort as the most crucial attributes of prosthetic sockets, with 83.3\% emphasizing these as significant considerations in their prosthetic selection process. The survey participants, ranging in age from 20 to 69 years, predominantly consisted of individuals actively using prosthetic devices, with 95 out of 97 respondents having a prosthesis and 80 individuals regularly utilizing their prosthetic devices. These findings underscore the importance of addressing durability and comfort concerns in prosthetic design and development to meet the evolving needs and expectations of individuals with limb loss. 

There are a lot of challenges associated with using prostheses in everyday life. Traditional prosthetic devices commonly use mechanical approaches, which can be inconvenient and fail to provide the fine motor control needed for more intuitive and accurate activities. Because of this, users frequently express unhappiness with the prostheses since they find them difficult to use and have limited usefulness.

BCI provides a significant opportunity to improve the control of prosthetics by utilizing brain impulses directly for operation. Nevertheless, existing BCI techniques encounter substantial constraints, as previously emphasized. They frequently struggle to accurately decipher the intricate patterns of motor imagery and adjust effectively to users' ever-changing cognitive states and motor intents. This constraint leads to a deficiency in accuracy and naturalness in controlling prosthetics, rendering it challenging for users to execute actions that need delicate motor abilities, such as grasping or moving things. Therefore, there is a critical need for novel methodologies to analyze and acquire knowledge from EEG data dynamically. By enhancing the flexibility and responsiveness of  BCI, these technologies can greatly improve prostheses' performance, providing users with a more natural and efficient means of controlling their mechanical limbs. Not only would this enhance the quality of life for persons with amputations, but it would also advance the limits of present prosthetic technology.

EEG is a cornerstone technology in BCI, leveraging the brain's electrical activity to interpret user intentions. EEG-based BCIs primarily utilize motor imagery (MI) signals generated when a user imagines a movement without actual physical execution. These signals are pivotal for applications such as prosthetic control but are notoriously subtle and subject to significant inter- and intra-individual variability. Accurately capturing and decoding these signals is still a difficult task due to the low ratio of signal-to-noise and the inherent variability of EEG data (\cite{degirmenci2023statistically}\cite{singh2021comprehensive}).

The capture of motor imagery signals by EEG-based BCIs is highly dependent on the examination of particular EEG frequency bands, including delta (1-4 Hz), theta (4-8 Hz), alpha (8-12 Hz), beta (12-25 Hz), and gamma (25-100 Hz). These frequency bands are very important for telling the difference between the start and end of motor imagery tasks, which helps in the more accurate sorting of motor intentions (\cite{vatrano2023assessing}; \cite{gurve2020trends}). By using these bands, especially through techniques such as spectrum filtering and spatial pattern recognition, BCIs can improve signal extraction and decoding accuracy, resulting in enhanced system responsiveness and user engagement. More details about different Waves are shown below:

\begin{itemize}
    \item \textbf{Delta Waves:} With frequencies ranging from 1 Hz to 4 Hz, delta waves are the slowest yet largest amplitude brainwaves. They predominantly occur during deep sleep stages and are more common in infants and young children.
    
    \item \textbf{Theta Waves:} Theta waves, oscillating between 4 Hz and 8 Hz, are associated with activities involving processing and learning. These waves become more evident when tackling complex tasks or engaging in significant mental effort. They can be detected across the brain's cortex and are easily recordable from the scalp.
    
    \item \textbf{Alpha Waves:} Alpha waves emerge within the 8 Hz to 12 Hz frequency range and are typically observed when an individual is in a state of relaxation. These waves are more pronounced during tasks that involve motor skills or memory, especially with closed eyes, and decrease during active mental or physical tasks.
    
    \item \textbf{Beta Waves:} Occurring in the frequency band of 12 Hz to 25 Hz, beta waves are linked to states of alertness and activity, such as concentrated thinking or anxiety. These waves indicate a high level of engagement and mental activity.
    
    \item \textbf{Gamma Waves:} Characterized by their high frequency above 25 Hz, gamma waves are associated with advanced cognitive functions like deep meditation or feelings of joy. These are often observed in individuals with extensive meditation experience, like monks.
\end{itemize}

Researchers in EEG-based BCIs have developed various methodologies to enhance these systems' adaptability and accuracy. One fundamental approach is the improvement of signal acquisition techniques to increase the quality and reliability of EEG data. This includes high-density EEG caps and advanced filtering technologies that reduce noise and increase the signal-to-noise ratio. Additionally, machine learning algorithms have become central to BCI development and have been employed to decode and classify EEG signals more effectively. These algorithms range from traditional linear classifiers to more complex models like support vector machines (SVM) and deep learning networks capable of handling EEG data's high dimensionality and variability (\cite{lotte2018review}). Several innovative approaches to address the specific challenges of real-time adaptability and motor imagery classification have been introduced in the last few decades. One widely used method is the common spatial pattern (CSP) filtering (\cite{alturki2021common}), which enhances the signal contrast between different motor imagery tasks. The CSP method makes it easier to tell the difference between EEG signals related to different motor imagery tasks by computing spatial filters that increase the variance for one class while decreasing it for the other. This method is particularly effective in applications where users need to control devices through distinct mental commands, and its effectiveness is enhanced when combined with machine-learning models that can adapt these filters in real-time. Another critical advancement is the integration of deep learning architectures such as convolutional neural networks (CNNs) (\cite{bang2021spatio}) and long short-term memory networks (LSTMs) (\cite{amin2021attention}). These models extract and learn features from complex, high-dimensional datasets like EEG signals. By training on large datasets, they can learn to recognise patterns associated with specific motor imagery tasks and adapt to changes in brain activity over time, significantly improving accuracy and responsiveness. Combining these methods has led to developing more sophisticated and adaptive BCIs that can better meet users' needs in real-time environments. This dynamic approach is at the heart of the latest advancements in the field, allowing for a more intuitive and effective interaction between the human brain and computer systems.

Researchers are increasingly utilising hybrid ML and DL methods to enhance performance (\cite{khademi2022transfer}). These hybrid approaches utilise the strengths of multiple algorithmic techniques, improving the accuracy and reliability of BCIs in decoding complex neural signals. Hybrid ML/DL models often combine traditional machine learning algorithms with advanced deep learning architectures to exploit structured feature engineering and automatic feature extraction capabilities. For instance, integrating support vector machines (SVM) with convolutional neural networks (CNNs) harnesses SVM's classification efficiency and CNN's capacity to extract hierarchical features from raw EEG data (\cite{saidi2021novel}). This combination enhances the ability to detect and classify subtle patterns in EEG signals with high accuracy, which is crucial for motor imagery recognition tasks \cite{schirrmeister2017deep}. Ensemble methods also play a crucial role, aggregating predictions from several deep learning models to improve predictive performance. By combining the outputs of different models trained on the same dataset, such as recurrent neural networks (RNNs) and CNNs, the ensemble reduces the risk of overfitting and increases the robustness of the system, essential for dynamic BCI applications \cite{cecotti2010convolutional}.
Furthermore, using transfer learning techniques, where a model developed for one task is repurposed for another, addresses the challenge of scarce labelled data in BCI research. This method allows for the utilisation of pre-trained networks on extensive datasets, which are then fine-tuned to specific BCI tasks, accelerating the training process and enhancing model performance \cite{dehghani2021deep}. These hybrid methods, combining multiple machine learning paradigms and leveraging deep learning  innovations, are instrumental in advancing the field of BCIs, offering more reliable, accurate, and user-adaptive systems.

Incorporating Natural Language Processing (NLP) models into EEG-based BCIs (\cite{hollenstein2021decoding}), especially those utilising reinforcement learning (RL) (\cite{xu2021accelerating}), represents a novel and promising direction in neurotechnology research. This approach allows BCI to interpret neural signals in ways that mimic human language processing, potentially transforming the intuitiveness and functionality of these systems. NLP-based models in BCIs are designed to analyze and interpret neural signals as if they were components of language. This involves the decomposition of signal patterns into features that resemble linguistic elements, which can be processed for intentions, emotional states, or specific commands. The application of NLP techniques such as semantic analysis, syntax parsing, and topic modelling can help in understanding the contextual and structural aspects of neural signals, making BCIs more natural for users to interact with.
Incorporating reinforcement learning into this framework enhances the adaptability of BCIs. Reinforcement learning algorithms learn optimal actions through trial-and-error interactions with a dynamic environment, adjusting their strategies based on feedback to maximise a reward function. In the context of NLP-based BCIs, RL can be used to fine-tune the interpretation of neural signals, continuously improving the system’s accuracy and responsiveness. For example, an RL algorithm could adjust parameters in real-time to better align the BCI’s output with the user’s intended commands based on the success or failure of previous interactions \cite{sutton2018reinforcement}.
A practical implementation of this is seen in systems where NLP and RL are used to refine decision-making processes, adapting to the user's specific neural patterns and preferences over time. These systems can learn, for instance, to recognise and differentiate between a broader array of mental commands by analysing the nuances in EEG signals as if they were parsing sentences in a spoken language. This results in a more accurate and user-tailored interface, which can be particularly beneficial in assistive technologies for individuals with mobility or speech impairments.

In this study, we introduce the Neuroassist model, a novel approach designed to transform EEG data analysis for prosthetic control. Traditional methods in this field often struggle with adaptability and responsiveness, which are critical for users with varying cognitive states and motor intentions. Neuroassist addresses these limitations by incorporating advanced techniques specifically chosen to optimize various facets of EEG data processing. For detailed feature extraction, we utilize the BERT base model to handle numerical EEG data. This adaptation allows the Neuroassist to discern subtle and complex patterns in EEG signals that are indicative of motor intentions. Additionally, Neuroassist integrates LSTM networks, which excel in managing temporal dynamics within sequential data. This integration enables the model to track EEG sequences over time, effectively capturing temporal dependencies essential for accurately predicting motor intentions. To facilitate efficient, low-power computation suitable for real-time applications, Neuroassist also employs Spiking Neural Networks (SNNs). These networks mimic biological neural dynamics, providing a robust framework for processing EEG data with minimal energy consumption.

Further enhancing its capabilities, Neuroassist incorporates Deep Q-Networks (DQN) to optimize decision-making processes. With DQN, the model may make use of feedback mechanisms that allow it to continuously adapt to the user's changing requirements and intents in real-time. Additionally, the model employs a preprocessing methodology that utilizes the CSP method for MI classification in an OVR way. This preprocessing not only keeps the temporal integrity of EEG signals but also makes classification results much better by pulling out distinguishing features. Our experimental evaluation of benchmark EEG motor imagery datasets confirms that Neuroassist achieves state-of-the-art performance. This underscores its considerable potential to advance assistive technology, especially in enhancing the adaptability and responsiveness of  BCI systems for users with diverse cognitive states and motor intentions.

\section{Literature Review}

The recent advancements in EEG-based MI research have showcased significant progress across various fronts, encompassing dataset collection, model development, and performance evaluation. This burgeoning field is primarily motivated by the imperative for robust BCIs capable of aiding rehabilitation and facilitating communication for individuals with disabilities. In parallel, recent studies in prosthetic control have observed a discernible shift towards the integration of ML, DL, and advanced AI techniques to surmount the constraints of traditional methodologies. These novel approaches harness sophisticated algorithms for feature extraction, temporal dynamics management, and real-time decision-making. For example, ML and DL models are increasingly being leveraged to extract intricate features from physiological signals like EEG data, thereby enabling more precise analysis and interpretation. Furthermore, advanced AI methodologies, including recurrent neural networks such as LSTM and spiking neural networks, are being deployed to enhance the adaptability and responsiveness of prosthetic devices, especially in real-time scenarios. These progressive developments underscore the transformative potential of incorporating state-of-the-art technologies into assistive devices, ultimately bolstering their efficacy in catering to the diverse requirements of individuals with disabilities. The comprehensive overview is covered in the following subsections:

\subsection{EEG Recordings}

\begin{figure}[htbp]
    \centering
    \includegraphics[width=0.8\textwidth]{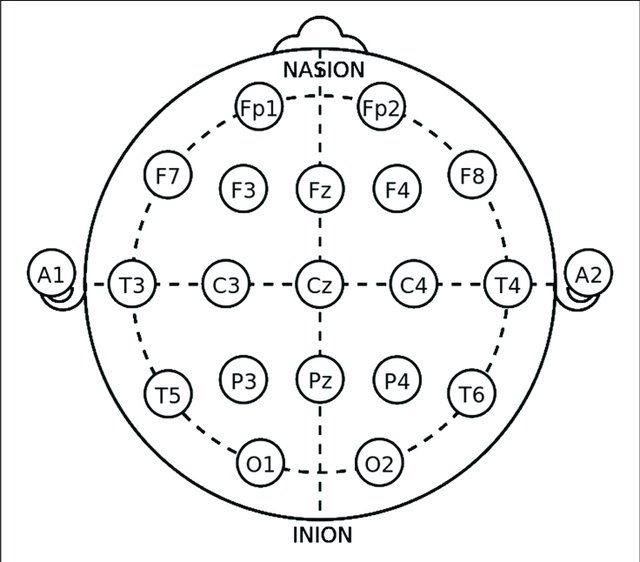}
    \caption{The international standard for EEG electrode placement is known as the 10-20 system. (\cite{rojas2018study})}
    \label{fig:electrode_position}
\end{figure}

The EEG tracks human physiological responses by recording electrical signals from cortical neurons during synaptic activity (\cite{sazgar2019overview}). These impulses are detected by conductive electrodes on the scalp using the 10-20 international system (\cite{8897723}). Electrodes capture a one-dimensional raw EEG data stream, capturing the brain's complex electrical activity via tissues to the scalp (\cite{sazgar2019overview}). See Figure \ref{fig:electrode_position} for electrode placement in the worldwide 10-20 system.

However, noise or interference from the eye, head, neck, or muscular movements might impair EEG results. The power cable and electrode movement of the recording equipment may also produce artifacts. Artifacts cause weak, non-stationary signals with poor signal-to-noise ratios, making classification and analysis difficult (\cite{sazgar2019overview}).

Despite problems, raw EEG data offers advantages over other brain imaging approaches, including price, mobility, and non-invasiveness (\cite{8698218}). It is used to screen and diagnose neurological problems, including epilepsy, brain tumours, Alzheimer's disease, and sleep disorders, by concentrating on wave patterns like fast spikes or slow waves. EEG signals have amplitudes in microvolts and frequency bands (delta, theta, alpha, beta, and gamma) that correlate to distinct brain activity types (\cite{8496885}). It is also worth mentioning the mu band, which represents alpha frequencies in the sensorimotor cortex, and the gamma band, which is weak on the scalp and can only be picked up by electrodes inside the brain (\cite{padfield2019eeg}).

Mental state greatly affects EEG signal complexity. Research suggests that automated systems are more accurate in classifying signals at the start of a session than at the conclusion (\cite{padfield2019eeg}). Therefore, EEG data must be carefully collected and processed to construct and evaluate machine learning models for this data.

\subsection{Dynamic AI Framework and Real-Time Synaptic Weight Adjustments}

Our NeuroAssist approach is a pioneering prosthetic control development combining an AI system with a dynamic and adaptable operational framework. This starkly contrasts the static models that have typically been employed in this industry. The AI system created for NeuroAssist is designed to adjust its synaptic weights in real-time, as explained in \cite{michaelis2022brian2loihi}. The ability to adapt and respond quickly to specific motor intentions detected from electroencephalogram (EEG) signals is crucial. This allows for immediate modifications to the architecture of the system, which is a significant departure from conventional brain-computer interface (BCI) technologies \cite{wang2012nonnegative}. The system's capacity to adapt in real-time allows it to constantly update and refine its performance, maintaining consistent accuracy in response to the changing patterns of brain activity.

Moreover, the ever-changing characteristics of an AI system indicate a significant change in how AI is incorporated into human neurological processes. The NeuroAssist project improves the efficiency of prosthetic devices by facilitating a smoother integration between artificial intelligence and brain signals. Moreover, this technology has novel opportunities for its utilization in other domains, such as medical diagnostics, rehabilitative treatment, and other assistive technologies. The AI's ability to quickly adjust to and engage with intricate cerebral processes in real time highlights its potential to transform human-machine interfaces.

\subsection{Feature Extraction methods in EEG}

This innovative effort uses an AI system to precisely extract characteristics from electroencephalogram (EEG) data. This approach analyzes signals from numerous frequency bands, such as gamma, mu, epsilon, and zeta, over many channels for multidimensional interpretation \cite{hummos2022thalamus}.

In this analytical approach, numerous crucial parameters are precisely calculated. Several metrics can be used to analyze EEG signals, such as median frequency, spectral entropy, and amplitude variability. The median frequency indicates the central value of the frequency distribution, while spectral entropy reveals the level of disorder or complexity within frequency components \cite{umichprosthetic}.

The system assesses signal deviation index, highlighting differences between expected and actual values \cite{michaelis2022brian2loihi}. Additionally, data asymmetry analysis enhances comprehension of EEG data skewness. The Hurst Exponent assessment provides information on distribution kurtosis, skewness, anomalies, and non-conforming data points \cite{tamuprosthesis}.

These carefully determined metrics are synthesized across many channels to offer a complete EEG data overview \cite{iparm}. This detailed feature extraction allows the AI system's neural network to categorize and identify motor intents with precision and efficiency.

These different variables from several sources are integrated and aggregated to provide a strong and thorough dataset for AI system learning algorithms. With this dataset, computers can make accurate predictions and judgments. This detailed feature extraction technique gives the AI system a comprehensive grasp of EEG data, allowing it to handle a range of inputs and making it a leading neural network-based analysis and prediction tool.

\subsubsection{Advanced Interpretation of Complex Neural Patterns}

Its state-of-the-art capabilities are demonstrated by the AI system's skill in understanding intricate brain combinations \cite{nptl}. The system shows remarkable proficiency in processing and comprehending complex EEG data, a difficulty for traditional analytical models, using advanced algorithms, and tapping into the dynamic principle of brain plasticity. The development of more accurate and responsive prosthetic controls is dependent on this skill, which also paves the way for novel approaches to brain-computer interfaces.

By going above and beyond using conventional data processing techniques, the system provides profound understanding of the complex neurological processes at action in the human brain. A major step forward in AI, the system's greater interpretative capabilities shows how it can change the way we interact with and comprehend the intricate workings of the human brain. This development emphasizes the critical importance of sophisticated AI in expanding our understanding of the brain, which in turn opens up new avenues for use in neurotechnology and medical research.

\subsection{Neuromorphic Cognitive Computing Approach }

Leading the way in artificial intelligence advancements, neuromorphic computing creates new paths in cognitive computing by fusing components like Deep Learning (DL) and Spiking Neural Networks (SNNs) (\cite{luo2020eeg}). This method leverages the combination of DL algorithms for contextual decision-making and SNNs for processing time-sensitive data, as demonstrated in my work with the NeuroAssist project (\cite{gong2023spiking}). We use the Intel Loihi chip, a specialized neuromorphic hardware that specializes in deploying SNNs for real-time applications in sectors like robotics and autonomous systems, and our work is particularly influenced by neurobiological processes.

A paradigm shift in AI research is embodied by combining SNNs with DL techniques. We enable advanced BCI solutions that can respond to intricate environmental stimuli by combining these technologies \cite{kumarasinghe2020deep}. The area of neuromorphic computing combines the analytical capabilities of DL systems with the high-speed, low-power processing of SNNs to emulate the computational efficiency and flexibility of the human brain. This potent mix is the foundation of our sophisticated computational models, tuned for optimal performance in the Intel Loihi neuromorphic device.

SNNs are an essential part of neuromorphic computing because of their capacity to mimic biological neural activity through spiking processes \cite{yamazaki2022spiking}. Compared to conventional neural networks, these networks more effectively capture the temporal diversity of real-world signals by processing input dynamically using spikes. SNNs' innate ability to process data efficiently and with little energy use is in line with the objectives of creating strong, environmentally conscious, and sustainable AI technology. Moreover, SNNs' capacity to withstand noisy environments makes them more applicable in a variety of real-world situations where noise and data integrity might present serious obstacles.

\subsection{NLP Functionality}

As a key component of modern artificial intelligence frameworks, NLP enables machines to comprehend, produce, and parse human language \cite{khurana2023natural}. NLP covers a wide range of activities, from entity recognition and sentiment analysis to more sophisticated operations like text summarization and machine translation. NLP can extract meaningful information from unstructured text using advanced techniques like deep learning, neural networks, and statistical analysis. This allows for a range of applications, from sophisticated translation services to conversational bots \cite{alshemali2020improving}. The ongoing improvements in natural language processing (NLP) have sparked innovations in many fields, revolutionizing communication, information retrieval, and strategic decision-making.

NLP powers the intelligence underlying chatbots \cite{ayanouz2020smart} and virtual assistants in industries like customer support, providing prompt help and customized replies. It supports diagnostic and therapeutic techniques in the medical area by automating the extraction of important facts from clinical records \cite{roy2021application}. NLP technologies are used in the financial industry to analyze large amounts of news and social media material in order to anticipate market movements and assess investment risks \cite{hodorog2022machine}. Furthermore, NLP makes it easier to translate across languages in real time, which improves cross-cultural communication.

The use of NLP methods for analysing EEG data is becoming popular due to its ability to provide a novel understanding of the brain's electrical activity. \cite{hollenstein2021decoding}. NLP is used by researchers to interpret patterns seen in EEG data, improving the categorization of brain states and opening the door for more sophisticated brain-computer interfaces (BCIs). With the goal of revolutionizing healthcare, communication, and HCI, this transdisciplinary approach combines the dynamic data representation of EEG with the strong analytical capabilities of NLP \cite{adam2020deep}, \cite{cao2020review}.

Using NLP paradigms, EEG data is becoming more and more understood as a neural dialect that can be understood using text processing-specific techniques. The sequential nature of EEG data is handled by methods like RNNs and LSTM models, which capture crucial temporal signal patterns \cite{galassi2020attention}. Furthermore, the incorporation of transformer topologies and attention processes into EEG investigations captures complex spatial-temporal correlations, which is similar to the achievements shown in NLP tasks.

Moreover, domain adaptation strategies play a critical role in improving tasks such as emotion detection and mental state assessments by transferring knowledge from large textual datasets to EEG analysis \cite{deepa2021bidirectional}, \cite{yenduri2024gpt}. In addition to these model advancements, crucial NLP preprocessing methods like noise reduction and signal augmentation are tailored for EEG data, increasing the effectiveness of ensuing analysis.

One cutting-edge method to improve language processing powers is the integration of Spiking Neural Networks (SNNs) with a BERT-based NLP model in the NeuroAssist BCI framework. With its improved ability to simulate cognitive processes, this hybrid system holds great promise for breaking down linguistic barriers and improving communication technology. The NeuroAssist project is combining NLP with SNNs to improve AI's operational acuity and its ability to more precisely mimic human cognitive functions, which is a major advancement in the evolution of intelligent systems.

\subsection{Classification approaches}

ML and DL architectures have greatly benefited from recent advances in artificial intelligence when it comes to the categorization of motor imagery (MI). With remarkable accuracy, these models are able to differentiate between different MI tasks by utilizing the temporal dynamics and spatial patterns that are inherent in EEG data.

The EEG-inception model is a novel CNN architecture that uses the inception-time network as its foundation. Zhang et al. \cite{zhang2021eeg} presented it, processing raw EEG data end-to-end and including a special data augmentation approach to improve classification accuracy by around 3\%. By separating class-independent information, Yang et al. \cite{yang2021novel} created a decoding method for MI EEG that greatly improves the model's accuracy.

In order to extract various characteristics and recognize temporal patterns from EEG signals, Schirrmeister et al. \cite{schirrmeister2017deep} built shallow and deep CNN architectures. Insufficient training data resulted in lower performance of the deep CNN model, despite its promise. It was found that various CNN layers may abstract different parts of the EEG data.

Moreover, Ma et al. \cite{ma2018improving} used BiLSTM in conjunction with LSTM to evaluate temporal and spatial features of the eegmmidb dataset, improving accuracy by 8.25\% in comparison to traditional techniques. In order to efficiently collect both spatial and temporal information at the same time, a hybrid DNN technique incorporating several neural networks has also been widely used in BCI.

Using STFT for CNN training, Xu et al. \cite{xu2019deep} transformed MI EEG signals into 2D time-frequency spectrum pictures, outperforming SVM and ANN classifiers in the process. Using CNNs for spatial feature extraction from MI EEG data and LSTMs for temporal feature extraction, Zhang et al. \cite{zhang2019novel} outperformed SVM classifiers in terms of accuracy.

She et al. \cite{she2019hierarchical} achieved a classification accuracy of 67.76\% by utilizing an ELM model with CSP for feature extraction on the BCIC IV 2b dataset. Ha et al. \cite{ha2019motor} achieved 78.44\% accuracy on the same dataset.

In their implementation of a Deep Belief Network (DBN) with Restricted Boltzmann Machines (RBM) for MI classification, Tang et al. \cite{tang2019semisupervised} achieved an accuracy of 83.55\% on time-series data from the BCIC IV 2b dataset. In order to handle one-dimensional spatial information, Wang et al. \cite{wang2018lstm} used LSTM networks on the BCIC IV 2a dataset, attaining 71\% accuracy.

Applying the GRU with ReLU and Tanh activation functions to the BCIC IV 2a and 2b datasets, Luo et al. \cite{luo2018exploring} achieved accuracies of 73.56\% and 82\%, respectively. Finally, for MI classification, Yang et al. \cite{yang2018deep} coupled CNN and LSTM networks on a private dataset, achieving an astounding 86.71\% accuracy.

Tabar and colleagues \cite{tabar2016novel} investigated a new CNN and sparse autoencoder (SAE) combination for MI task classification using BCIC II 3 and IV 2b datasets, yielding accuracies of 77.6\% and 90\%, respectively.

\section{Methodology}
\subsection{Data Description}
\subsubsection{GigaScience dataset}

This study utilizes the publicly available GigaScience dataset with our proposed approach. This multimodal dataset is designed for brain-computer interface (BCI) research focusing on upper-extremity movements \cite{nemati2022feature}. It was compiled from electroencephalogram (EEG) recordings of 25 healthy subjects across multiple sessions to capture various intuitive movement tasks associated with the arms, hands, and wrists. The dataset comprises 82,500 trials across all subjects, averaging approximately 3,300 trials per subject.

To provide thorough coverage of the scalp and capture essential brain signals for motor and sensory processing, the EEG data were collected using a 60-channel setup based on the worldwide 10–20 position system. The dataset also includes recordings from seven-channel EMG and four-channel EOG, enriching the dataset's multimodal nature. This multimodal approach facilitates a comprehensive understanding of neural and muscular activities during motor tasks and supports the development of more sophisticated and intuitive BCIs.

This dataset obtained 108 EEG files in Matlab format. The PSD was computed for selected channels. The extracted features were concatenated to form a dataset with a structure of (399,) and normalized using MinMax normalization to maintain consistency across training and testing phases.

\subsubsection{BCI-IV2a dataset}
In this study, the BCI Competition 2008 dataset 2a is used to validate our neuroassist approach. This dataset of nine people visualizing four body parts is meant for decoding motor imagery tasks from EEG data \cite{altuwaijri2022electroencephalogram}.

Structure of the dataset and recording protocol: two sessions with six runs for each of the nine individuals. The four motor imagery tasks—left hand, right hand, feet, and tongue—are randomly assigned 48 trials per run. With 576 trials per person and 288 trials per session, the total number of trials in the sample is 5,184.

Recording and Preprocessing: EEG data were gathered at 250 Hz. Data quality was ensured with a 0.5–100 Hz band-pass filter. For artifact processing, three monopolar EOG channels were recorded. To analyze EOG artifacts, each session began with five minutes of recording with eyes open, closed, and eye movements.

Data Format and Use: The GDF stores one file per subject every session. Class labels are included in all trial training data but not in classifier testing data. The BCI competition evaluated continuous classification output for each sample, including artifacts, using accuracy time courses and kappa coefficient.

 Feature Extraction and Classification: To extract and classify features, we analyzed signals from the first session of each participant, except for Participant 4, due to inadequate data. ICA was repeated to reduce artifacts. One-vs-the-rest CSP was used to extract discriminative features for each motor imaging challenge. This required translating EEG signals into a "csp space," retaining the time points per CSP spatial component, and stacking statistical measurements and PSD estimations of the alpha and beta bands across components to build a 2D feature array. The suggested model was trained using extracted characteristics.

\subsection{Feature Extraction Methods}
\textbf{Welch’s Method:}
Welch's approach is a reliable methodology used to estimate a signal's power spectral density (PSD). This is an important aspect of signal processing as it helps assess how power is distributed across different signal frequency components. This method improves the periodogram methodology by dividing the time series data into segments, using window functions on each segment to reduce bias, and then averaging these adjusted periodograms to provide a reliable spectral density estimate. The primary equation of this approach (\cite{chiu2023quantifying}) is as follows:

\begin{equation}
P_k(f) = \left| \frac{1}{\sqrt{L}} \sum_{n=0}^{L-1} x_k[n] \cdot e^{-i 2\pi \frac{f n}{L}} \right|^2
\end{equation}

In this equation, $P_k(f)$ denotes the estimated power at frequency $f$, $x_k[n]$ is the $n$-th sample in the $k$-th data segment, $L$ represents the length of each segment. The exponential term embodies the Fourier transform, converting time-domain data points into their frequency-domain counterparts. Squaring the Fourier transform's modulus yields the segment's power spectrum, averaged over all segments to enhance estimate accuracy and reduce variance.

This technique considerably improves spectrum estimation by reducing variance and bias compared to the basic periodogram, rendering it especially advantageous for crucial applications such as EEG analysis \cite{keil2022recommendations}. It performs exceptionally well in circumstances with high noise levels or when minimal data is available. It offers flexibility through different window functions and segment overlaps, allowing for customized spectral analysis.

\begin{equation}
P_{xx}(f) = \text{median}\{P_0(f), P_1(f), \ldots, P_{K-1}(f)\}
\end{equation}

In this equation, $P_{xx}(f)$ represents the estimated power spectral density at frequency $f$. It is calculated as the median of the individual power spectral density estimates $P_0(f), P_1(f), \ldots, P_{K-1}(f)$. Taking the median helps to reduce the impact of outliers and noise present in the individual estimates, providing a more robust and reliable estimation of the power spectral density at each frequency.

\textbf{Kurtosis:}
The statistical measure of kurtosis provides information about the dispersion of values in a dataset. This tool can better understand the presence or absence of outliers in the tails of a normal distribution. There are several ways to compute it, but a popular one involves solving for the dataset's fourth central moment and then normalizing it by raising it to the standard deviation's power, making it independent of scale. The given formula estimates the population kurtosis by representing the sample kurtosis \cite{chakraborty2021epilepsy}. 

\begin{equation}
\text{Kurtosis} = \frac{n(n + 1)}{(n - 1)(n - 2)(n - 3)} \sum_{i=1}^{n} \left( \frac{x_i - \bar{x}}{s} \right)^4 - \frac{3(n - 1)^2}{(n - 2)(n - 3)}
\end{equation}

The parameters used are the number of observations ($ n $), individual observations ($ x_i $), sample mean ($ \bar{x} $), and sample standard deviation ($ s $). This equation calculates the fourth central moment of the sample. It standardizes the moment by dividing it by the fourth power of the standard deviation.

\textbf{Root Mean Square (RMS):}

RMS is a popular statistical metric that shows the magnitude of a collection of values. Signal processing and data analysis often measure fluctuating signal effective amplitude. RMS is the square root of the average squared values of each data point. The RMS equation is mathematically:

\begin{equation}
\text{RMS} = \sqrt{\frac{1}{n} \sum_{i=1}^{n} x_i^2}
\end{equation}

In this scenario, the symbol \( \text{RMS} \) indicates the Root Mean Square value, \( n \) stands for the total number of data points, and \( x_i \) refers to the individual data points within the collection. The equation calculates the root mean square (RMS) value by squaring each data point (\( x_i^2 \)), averaging them across all data points (\( \frac{1}{n} \sum_{i=1}^{n} x_i^2 \)), and then taking the square root. This procedure accurately considers both positive and negative values, obtaining a measure of the overall size of the collection.

\textbf{Skewness:}
Skewness measures distribution asymmetry around its mean via statistical means. It shows if data is left or right of the mean. Positive and negative skewness imply a longer right and left tail, respectively. For skewness, the departure of each data point from the mean is raised to the power of 3, normalized by the standard deviation, and averaged. Skewness equation mathematically looks like:

\begin{equation}
\text{Skewness} = \frac{n}{(n - 1)(n - 2)} \sum_{i=1}^{n} \left( \frac{x_i - \bar{x}}{s} \right)^3
\end{equation}

Here, \( \text{Skewness} \) represents the skewness value, \( n \) represents the total number of data points, \( x_i \) represents individual data points within the dataset, \( \bar{x} \) represents the mean of the dataset, and \( s \) represents the standard deviation of the dataset. The equation normalizes the variation of each data point from the mean by dividing it by the standard deviation, which guarantees scale-invariance. To calculate the distribution's skewness, the deviations are first cubed, then added together, and then divided by \( n \) times \((n - 1)(n - 2)\) to get the average cubed deviation.

\textbf{Absolute Difference:}
Absolute Difference is a mathematical metric used to quantify the disparity between two sets of data. The algorithm computes the total of the absolute disparities between matching members of two sets. This measure is very valuable in the fields of image processing, optimization issues, and data comparison activities. The Absolute Difference between two sets of values \(x_i\) and \(y_i\) is calculated by adding up the absolute differences between each corresponding element, as shown in the equation:

\begin{equation}
\text{Absolute Difference} = \sum_{i=1}^{n} |x_i - y_i|
\end{equation}

Here, \( n \) is the total number of items in the sets, \( | \cdot | \) is the absolute value function, and \( \text{Absolute Difference} \) is the resulting measure of disagreement. The size of the difference between the \(i\)-th elements of the two sets is represented by each absolute difference \( |x_i - y_i| \) in the equation. An overall assessment of the sets' differences may be obtained by adding these absolute differences across all components.

\textbf{OVR and CSP for Multiclass MI-EEG Classification:}

CSP is a very effective approach to extracting features from multichannel EEG in BCI systems. The objective is to optimize the variance of EEG signals belonging to a certain class while reducing the variance of signals from other classes. In categorizing numerous MI tasks using EEG, the CSP method is often utilized in conjunction with the OneVsRest (OVR) technique. The OVR technique entails training a binary classifier for each class compared to the remaining classes.

\begin{equation}
w^*_i = \underset{{w_i}}{\text{arg max}} \frac{{w_i^T \Sigma_{\text{class} \; i} w_i}}{{w_i^T \Sigma_{\text{rest}} w_i}}
\end{equation}

In this case, \(w_i\) refers to the spatial filter for class \(i\), whereas \(w_i^T\) represents the transpose of \(w_i\). The fraction's numerator reflects the variance of class \(i\), whereas the denominator represents the variance of the other classes.

\subsection{Actor-Critic Framework}

This section discusses the development and structure of the actor-critic method, as highlighted in \cite{haarnoja2018soft}. This method involves alternating updates across $T$ steps. At each step $t$, a policy $\pi_t$ is defined by the parameter $\theta_t$. The critique improves its value function estimate with parameter $w_t$ using on-policy data obtained by applying $\pi_t$ to the base model over a $K$-length horizon. Essentially, the data from the policy is crucial for updating the critic and actor components.

Significant outcomes are achieved in the RNAC method, particularly under the LFA. Here, the critic is updated through a resilient linear-TD method, while the actor applies a robust QNPG update, as outlined in Algorithm \ref{fig:algorithm1}.

\textbf{Theorem 1}
Implemented within Algorithm \ref{fig:algorithm1}, the robust natural actor-critic (RNAC) framework makes use of linear function approximation and operates under either data-driven sampling (DS) or importance sampling with model perturbation (IPM) uncertainties. The framework conducts updates through a robust linear temporal difference (RLTD) mechanism for the critic and a Robust Q-Network Policy Gradient (RQNPG) method for the actor, with progressively increasing step sizes $\eta_t$ \cite{zhou2024natural}.
\[
E[V^{\pi^*}(\rho) - V^{\pi^T}(\rho)] = O(e^{-T}) + O(\epsilon_{\text{stat}}) + O(\epsilon_{\text{bias}})
\]
In this theorem, the gap towards optimality, $E[V^{\pi^*}(\rho) - V^{\pi^T}(\rho)]$, breaks down into three key components. The exponential decline with respect to the number of iterations $T$ as $O(e^{-T})$ illustrates the rapid convergence. Statistical discrepancies, indicated by $\epsilon_{\text{stat}} = \tilde{O}(\sqrt{1/N} + \sqrt{1/K})$, stem from the sizes of the sample sets $N$ and $K$, which inform the updates to both the critic and actor. The term $\epsilon_{\text{bias}}$ highlights the limitations in the approximation capabilities of the utilized value function and policy class.
Overlooking $\epsilon_{\text{bias}}$, we determine that to reach an $\epsilon$-robust optimal outcome, the required sample complexity is roughly $\tilde{O}(1/\epsilon^2)$. This demonstrates efficiency comparable to a tabular approach. The introduction of geometrically scaled step sizes in the RNAC algorithm results in a larger coefficient in the convergence expression.

\textbf{Theorem 2}
Using a fixed step size $\eta_t = \eta$ and under the identical circumstances as Theorem 1, the RNAC exhibits the following behaviour in terms of convergence (\cite{zhou2024natural}): 

\[
E[V^{\pi^*}(\rho) - \frac{1}{T} \sum_{t=0}^{T-1} V^{\pi_t}(\rho)] = O\left(\frac{1}{T}\right) + O(\epsilon_{\text{stat}}) + O(\epsilon_{\text{bias}})
\]

This results in an estimated sample complexity of $\tilde{O}(1/\epsilon^3)$.
While this theorem indicates a reduced rate of optimization when RNAC operates with a uniform step size, such step sizes are often preferred in practical applications. Additionally, this approach can be extended to a broader range of policy classes, predicting an optimization rate of $O(1/\sqrt{T})$ and a sample complexity of $\tilde{O}(1/\epsilon^4)$ (\cite{zhou2024natural}).

\setcounter{figure}{0}
\begin{figure}[htbp]
    \centering
    \includegraphics[width=1\textwidth]{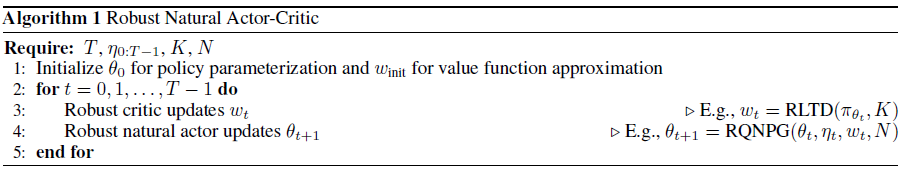}
    \refstepcounter{figure}
    \label{fig:algorithm1}
\end{figure}

\subsubsection{Robust Critic Implementation}

The robust critique aims to estimate the robust value function from nominal model samples. Previous actor-critic analyses in canonical RL often teach the Q function to the critic\cite{wang2022policy}. Proximal policy optimization and other practical on-policy methods target the V function to increase training efficiency.

The resilient linear temporal difference (RLTD) method Algorithm \ref{fig:algorithm2} combines an empirically Bellman operator with the classic linear-TD method. Iterative sampling and updating are done using this algorithm. In Algorithm \ref{fig:algorithm2}, the sampling process uses approaches like double sampling and Importance Sampling with Model Perturbation (IPM) based on the uncertainty set. For the linear value function approximation \cite{chen2022finite}\cite{li2022first}.

\textbf{Assumption 1}

The network of Markov chains $\{s_k\}$, generated by implementing the policy $\pi$ in the theoretical model $p^\circ$, is geometrically ergodic and converges to a unique stationary distribution $\nu_\pi$ for each given policy $\pi$. The updating mechanism effectively targets the reduction of the MSPRBE, as detailed in (\cite{zhou2024natural}).

\[
\text{MSPRBE}_\pi(w) = \lVert \Pi_\pi(T^\pi P Vw) - Vw \rVert_{\nu_\pi}^2
\]

where the weighted estimates matrix is \(\Pi_\pi = \Psi(\Psi^\top D_\pi \Psi)^{-1}\Psi^\top D_\pi\), the \(\nu_\pi\)-weighted norm is \(\lVert V \rVert_{\nu_\pi}\), and \(D_\pi = \text{diag}(\nu_\pi) \in \mathbb{R}^{S\times S}\). The projected equation \(Vw = \Pi_\pi T^\pi P Vw\) has a unique solution, is the minimizer of MSPRBE\(_\pi(w)\), represented by \(w_\pi\). As a result, since the empirical operator \(\hat{T}^\pi P\) is unbiased, RLTD is a stochastic approximation method (\cite{zhou2024natural}) with:

\[
\mathbb{E}\left[\left( \hat{T}^\pi P Vw - Vw \right) \psi(s) \right] = \Psi^\top D_\pi (T^\pi P Vw - Vw)
\]

To achieve convergence to the optimal linear approximation $Vw_\pi$ using RLTD, it is necessary that the projected robust bellman operator $\Pi_\pi T^\pi P$ behave as a contraction mapping with a contraction factor $\beta < 1$.

\textbf{Definition 2}

The operator $\Pi_\pi T^\pi P$ qualifies as a $\beta$-contraction with respect to the norm $\lVert \cdot \rVert_{\nu_\pi}$ if it satisfies the condition $\lVert \Pi_\pi T^\pi P V - \Pi_\pi T^\pi P V' \rVert_{\nu_\pi} \leq \beta \lVert V - V' \rVert_{\nu_\pi}$. In contrast to linear TD methods for MDPs, Assumption 2 shows that $\Pi_\pi T^\pi P$ is actually contracting.

\textbf{Assumption 2}

There is a constant $\beta \in (0, 1)$ such that the inequality $\gamma_{p,s,a}(s') \leq \beta p^\circ_{s,a}(s')$ holds across all states $s, s'$ in $\mathcal{S}$, for every action $a$ in $\mathcal{A}$, and for all probability distributions $p$ in $\mathcal{P}$.

\textbf{Proposition 2}

Assuming Assumption 2, $\Pi_\pi T^\pi P$ operates as a $\beta$-contraction for the norm $\lVert \cdot \rVert_{\nu_\pi}$. When $\delta$ is tiny, the uncertainty set $P$ with double-sampling violates Assumption 2, ensuring the contraction condition of $\Pi_\pi T^\pi P$. Using $m = 2$ in Equation (4), a $\delta$ less than $1 - \frac{\gamma}{2\gamma}$ is adequate.

However, selecting a small $\delta$ does not solve every set variation.

\textbf{Proposition 3}

In a geometrically mixed nominal model, the uncertainty set violates Assumption 2 for every $f$-divergence with a radius $\delta > 0$.

While Assumption 2 is widely recognized, it may not always be necessary. As illustrated in Equation (6), the Importance Sampling with Model Perturbation (IPM) uncertainty set aligns robustness with regularization, providing a clear formulation of the robust Bellman operator. The contraction properties of $\Pi_\pi T^\pi P$ for the IPM set can thus be evaluated independently of Assumption 2.

\textbf{Lemma 1}

Within the framework of the Importance Sampling with Model Perturbation (IPM) uncertainty set, if $\delta$ is less than $\lambda_{\text{min}}(\Psi^\top D_\pi \Psi)^{\frac{1-\gamma}{\gamma}}$, then $\Pi_\pi T^\pi P$ exhibits contraction properties in the norm $\lVert \cdot \rVert_{\nu_\pi}$.

In the context of robust Markov decision Processes (RMDP), contraction of $\Pi_\pi T^\pi P$ is ensured under both double sampling (DS) and IPM uncertainty sets, provided the radius $\delta$ is sufficiently small. Recent developments in the theory of Markovian stochastic approximations have led to the establishment of the first finite sample guarantees for Robust Linear Temporal Difference (RLTD):

\textbf{Theorem 3}

When RLTD is applied with step sizes $\alpha_k = \Theta(1/k)$, it achieves an expected convergence such that $E[\lVert w_K - w_\pi \rVert^2] = \tilde{O}\left(\frac{1}{K}\right)$.

\setcounter{figure}{1}
\begin{figure}[htbp]
    \centering
    \includegraphics[width=1\textwidth]{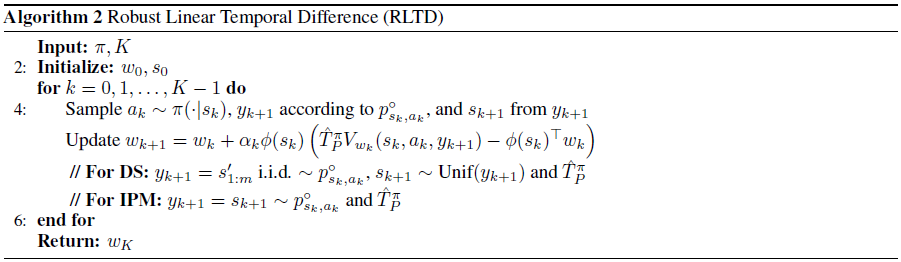}
    \refstepcounter{figure}
    \label{fig:algorithm2}
\end{figure}

\textbf{Robust Natural Actor}

In the domain of RL, strategies such as TRPO and PPO utilize the Kullback-Leibler (KL) divergence to ensure stable policy updates. These methodologies are based on the principles of the natural policy gradient (NPG), aimed at refining policy parameters $\theta$ by incorporating preconditioning to facilitate effective policy enhancements.

In the context of Robust Markov Decision Processes (RMDP), the policy $\pi_\theta$, parameterized by $\theta$ and inherently differentiable, is initially examined through the lens of the policy gradient theorem. Leveraging Rademacher's theorem, the robust value function $V^{\pi_\theta}_P(\rho) := J(\theta)$ is characterized as Lipschitz continuous with respect to $\theta$ and differentiable almost universally. Where $J$ is non-differentiable, we employ the notion of a Fréchet supergradient $\nabla J$, as defined by (\cite{zhou2024natural}):

\[
\limsup_{\theta' \to 0} \frac{J(\theta') - J(\theta) - \langle \nabla J(\theta), \theta' - \theta \rangle}{\lVert \theta' - \theta \rVert} \leq 0.
\]

This framework paves the way for the policy gradient theorem in robust RL, representing an extension of the classical RL policy gradient theorem:

\textbf{Lemma 2}

The gradient of the value function \( V^\pi(\rho) \) \cite{zhou2024natural} for a differentiable policy \(\pi = \pi_\theta\), parameterized by \(\theta\), is given by:
\[
\nabla_\theta V^\pi(\rho) = \frac{1}{1 - \gamma} \mathbb{E}_{s \sim d^\pi, \kappa^\pi_\rho} \left[ \mathbb{E}_{a \sim \pi_s} \left[ Q^\pi(s, a) \nabla_\theta \log \pi(a|s) \right] \right]
\]

where $\kappa^\pi_\rho$ represents the worst-case transition kernel under policy $\pi$.

For log-linear strategy, where $\pi_\theta(a|s) = \frac{\exp(\phi(s,a)^\top \theta)}{\sum_{a'} \exp(\phi(s,a')^\top \theta)}$ and $\phi(s, a) \in \mathbb{R}^d$ is the feature vector, Natural Policy Gradient (NPG) updates are performed as follows (\cite{zhou2024natural}):

\[
\theta \leftarrow \theta + \eta F_\rho(\theta)^+ \nabla_\theta V^{\pi_\theta}_{p^\circ}(\rho),
\]

with $F_\rho(\theta)^+$ being the moore-penrose-inverse of the fisher-information matrix $F_\rho(\theta) = \mathbb{E}_{(s,a) \sim d^{\pi_\theta,p^\circ}_\rho} [\nabla_\theta \log \pi_\theta(a|s) (\nabla_\theta \log \pi_\theta(a|s))^\top]$.

The update mechanism, $\theta_{t+1} = \text{RQNPG}(\theta_t, \eta_t, w_t, N)$, with $\zeta_n = \Theta(1/n)$, fosters policy improvement (\cite{zhou2024natural}). Theorem 4 claims:

\[
V^{\pi_{t+1}}(\rho) \geq V^{\pi_t}(\rho) + \frac{\text{KL}_d(\pi_{t+1},\kappa_{\pi_{t+1}}^\rho)(\pi_t, \pi_{t+1}) + \text{KL}_d(\pi_{t+1},\kappa_{\pi_{t+1}}^\rho)(\pi_{t+1}, \pi_t)}{(1 - \gamma)\eta_t} - \frac{\epsilon_t}{1 - \gamma},
\]

where $\text{KL}_\nu(\pi, \pi')$ measures the KL divergence and $\mathbb{E}[\epsilon_t] = \tilde{O}(\sqrt{\frac{1}{N}} + \sqrt{\frac{1}{K}}) + O(\epsilon_{\text{bias}})$, ensuring an approximate enhancement in policy value after each update step.

\subsubsection{Robust Natural Actor}

In Reinforcement Learning (RL), strategies such as Trust Region Policy Optimization (TRPO) and Proximal Policy Optimization (PPO) enhance policy parameters through preconditioning by the Kullback-Leibler (KL) divergence, \(\text{KL}(p, q) := \langle p, \log(p/q) \rangle\). These approaches are based on the natural policy gradient (NPG) method, prevalent in standard RL. Initially, we examine the policy gradient theorem for Robust Markov Decision Processes (RMDP), where the policy \(\pi_\theta\) is differentiably parameterized by \(\theta\) (\cite{zhou2024natural}).

The robust value function \( V_{\pi_\theta}^P(\rho) := J(\theta) \) is typically Lipschitz, hence differentiable almost everywhere according to Rademacher's theorem. In instances of non-differentiability, a Fréchet supergradient \(\nabla J\) exists, satisfying:
\[
\limsup_{\theta' \to \theta} \frac{J(\theta') - J(\theta) - \langle \nabla J(\theta), \theta' - \theta \rangle}{\|\theta' - \theta\|} \leq 0.
\]

The Q-Natural Policy Gradient (Q-NPG) (\cite{zhou2024natural}) updates the function by:
\[
\theta \leftarrow \theta + \eta u',
\]
where \(u'\) is determined by:
\[
u' = \arg\min_u \mathbb{E}_{(s,a) \sim d_{\pi, p^\circ} \rho \pi}[(Q_{\pi}^{p^\circ}(s, a) - u^\top \phi(s, a))^2],
\]
defining \(Q_u(s, a) := \phi(s, a)^\top u\) as a Q-value function approximation compatible with log-linear policies. \(Q_\pi\) guides the ascent direction per Lemma 2.

In robust RL, estimating \(Q_{\pi}^{p^\circ}\) from sampled trajectories is challenging. We use a critic's value function approximation \(V_w\) to estimate the probability of the robust Q-function:
\[
Q_w(s, a) = r(s, a) + \inf_{p \in \mathcal{P}_{s,a}} p^\top V_w,
\]
which aligns with \(Q_\pi\) when \(V_w = V_\pi\). The Robust Q-Natural Policy Gradient (RQNPG) iteratively refines this approximation through the critic's value function \(V_w\) and a robust Q-approximation \(Q_u\), detailed in Algorithm \ref{fig:algorithm3}. Sampling and update procedures follow the RLTD, Algorithm \ref{fig:algorithm2}) methodology. The update step is defined by:
\[
\theta_{t+1} = \text{RQNPG}(\theta_t, \eta_t, w_t, N),
\]
with \(\zeta_n = \Theta(1/n)\), guarantees policy improvement under robust approximation conditions.

\textbf{Approximate Policy Improvement}

\begin{theorem}[Approximate Policy Improvement]
For any \( t \geq 0 \), it is established that the value function \( V_{\pi_{t+1}}(\rho) \) after the update at time \( t+1 \) is at least as great as the value function \( V_{\pi_t}(\rho) \) at time \( t \), adjusted by the divergence between the policies at times \( t \) and \( t+1 \), and a term dependent on the step size \( \eta_t \) and the approximation error (\cite{zhou2024natural}):
\[
V_{\pi_{t+1}}(\rho) \geq V_{\pi_t}(\rho) + \frac{\eta_t}{1-\gamma} \left( \text{KL}_{d_{\pi_{t+1}, \kappa_{\pi_{t+1}} \rho}}(\pi_t, \pi_{t+1}) + \text{KL}_{d_{\pi_{t+1}, \kappa_{\pi_{t+1}} \rho}}(\pi_{t+1}, \pi_t) \right) - \frac{\epsilon_t}{1-\gamma},
\]
where \(\text{KL}_\nu(\pi, \pi') = \sum_{s} \nu(s) \text{KL}(\pi(\cdot|s), \pi'(\cdot|s)) \geq 0\) and \(E[\epsilon_t] = \tilde{O}\left(\sqrt{\frac{1}{N}} + \sqrt{\frac{1}{K}}\right) + O(\epsilon_{\text{bias}})\).
\end{theorem}

\setcounter{figure}{2}
\begin{figure}[htbp]
    \centering
    \includegraphics[width=1\textwidth]{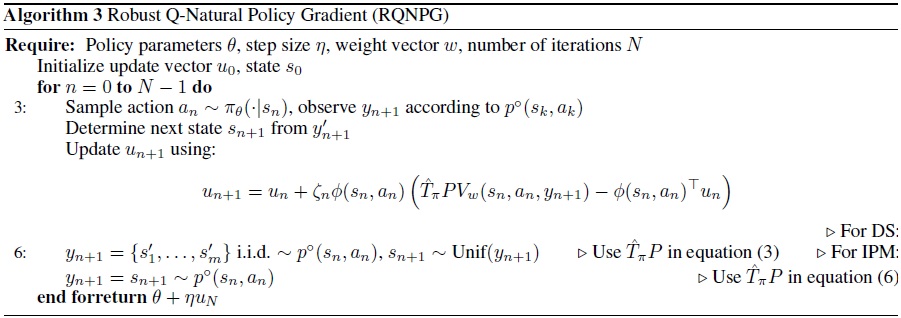}
    \refstepcounter{figure}
    \label{fig:algorithm3}
\end{figure}

\subsection{Developing Architecture}

\subsubsection{Bidirectional Encoder Representations from Transformers (BERT)}

BERT, an abbreviation for Bidirectional Encoder Representations from Transformers \cite{hollenstein2021decoding}, is a cutting-edge technique in Natural Language Processing (NLP). The introduction of BERT, created by Google researchers, was a major advancement as it demonstrated the ability to handle many NLP tasks using a single pre-trained model efficiently. BERT, unlike previous models, is designed to analyze the whole text sequence in a bidirectional way rather than focusing just on the beginning or end in order to predict the context. This allows it to catch more nuanced and contextually meaningful implications of words.

\textbf{Attention Mechanism:}
The multi-head self-attention mechanism is the foundation of the BERT model, as shown in the equation:
\[
\text{Attention}(Q, K, V) = \text{softmax}\left(\frac{QK^T}{\sqrt{d_k}}\right)V
\]

The variables $Q$, $K$, and $V$ correspond to the queries, keys, and values, respectively. The variable $d_k$ represents the dimension of the keys.

\textbf{Layer Normalization:}
Layer normalization, used to stabilize the learning process, is computed as:

\[
\text{LayerNorm}(x) = \gamma \left(\frac{x - \mu}{\sigma}\right) + \beta
\]

where $x$ is the input vector, $\mu$ and $\sigma$ are the mean and standard deviation of $x$, and $\gamma$ and $\beta$ are learnable parameters.

\textbf{Position-wise Feed-Forward Networks:}
Each layer of BERT includes a feed-forward neural network applied independently to each position:

\[
\text{FFN}(x) = \max(0, xW_1 + b_1)W_2 + b_2
\]

with $W_1, b_1, W_2, b_2$ being the weights and biases of the linear transformations.

\textbf{Masked Language Modeling (MLM):}
BERT is trained using the MLM objective, which randomly masks tokens and predicts their original vocabulary based on context. The loss function is usually cross-entropy.

\subsection{LSTM Architecture}
LSTM networks are a specific sort of RNN architecture designed to address the issue of the vanishing gradient problem, which may arise when conventional RNNs are used to handle lengthy data sequences. LSTMs are efficient for natural language processing, voice recognition, and sequence prediction because they can store information for a long time. Gates govern information transfer in LSTMs, preserving long-term memory \cite{wang2018lstm}. Gates include:

\begin{itemize}
    \item \textbf{}{Forget Gate:} Determines which components of the cell state should be eliminated.
  \item \textbf{Input Gate:} Determines the specific information to be stored in the cell state.
  \item \textbf{Output Gate:} Determines the subsequent concealed state and the desired output depending on the current cell state.
\end{itemize}

The behaviour of the LSTM unit may be characterized by the following equations, which dictate the state changes at each time step \(t\):

\textbf{Forget Gate:}
\begin{equation}
f_t = \sigma(W_f \cdot [h_{t-1}, x_t] + b_f)
\end{equation}

\textbf{Input Gate:}
\begin{equation}
i_t = \sigma(W_i \cdot [h_{t-1}, x_t] + b_i)
\end{equation}
\begin{equation}
\tilde{C}_t = \tanh(W_C \cdot [h_{t-1}, x_t] + b_C)
\end{equation}
where \(i_t\) is the output of the input gate, and \(\tilde{C}_t\) represents the candidate values for updating the cell state.

\textbf{Cell State Update:}
\begin{equation}
C_t = f_t \ast C_{t-1} + i_t \ast \tilde{C}_t
\end{equation}

By combining the prior state with the new candidate values, which are modulated by the forget gate and input gate outputs, this equation updates the cell state.

\textbf{Output Gate:}
\begin{equation}
o_t = \sigma(W_o \cdot [h_{t-1}, x_t] + b_o)
\end{equation}
\begin{equation}
h_t = o_t \ast \tanh(C_t)
\end{equation}

\subsection{Spiking Neural Networks}
Spiking neural networks (SNNs) \cite{antelis2020spiking} are a kind of neural computing that closely imitates biological neural networks. Contrary to conventional artificial neural networks, SNNs handle information in a temporal manner, where neurons communicate through distinct spikes. This technique mirrors the natural neuronal activity found in the brain.

SNNs integrate the temporal aspect into their functioning by having neurons activate when the inputs they receive exceed certain thresholds. This methodology enables SNNs to possibly function with more efficiency in terms of energy consumption and computing resources, particularly in situations where real-time processing is of utmost importance.

\textbf{Mathematical Model of Neuron Dynamics}

The dynamics of neurons within a spiking neural network (SNN) can be represented using differential equations. One frequently employed model is the leaky integrate-and-fire (LIF) model, which approximates the behaviour of biological neurons through an electrical circuit analogy.

\textbf{Membrane Potential Equation}

The evolution of the membrane potential \( V(t) \) in a neuron can be described by the following differential equation:

\begin{equation}
\tau \frac{dV}{dt} = -V(t) + RI(t)
\end{equation}

where \(\tau\) is the membrane time constant, \(R\) is the membrane resistance, and \(I(t)\) denotes the input current.

\textbf{Firing Condition and Reset Mechanism}

A neuron generates a spike when its membrane potential exceeds a specific threshold \( V_{\text{thresh}} \). After firing, the membrane potential is reset according to:

\begin{equation}
\text{if } V(t) \geq V_{\text{thresh}} \text{ then }
\begin{cases}
\text{emit spike} \\
V(t) \leftarrow V_{\text{reset}}
\end{cases}
\end{equation}

\textbf{Synaptic Inputs}

The input current resulting from synaptic inputs is typically modeled as a sum of weighted delta functions, representing the times at which spikes from other neurons are received:

\begin{equation}
I(t) = \sum_{i} W_i \cdot \delta(t - t_i)
\end{equation}

where \( W_i \) are the synaptic weights and \( \delta \) represents the Dirac delta function.

\textbf{Spike-Timing-Dependent Plasticity (STDP)}

STDP is a mechanism for adjusting synaptic weights based on the timing difference between pre-and post-synaptic spikes:

\begin{equation}
\Delta W = 
\begin{cases} 
A_+ e^{-\frac{\Delta t}{\tau_+}} & \text{if } \Delta t > 0 \\
-A_- e^{\frac{\Delta t}{\tau_-}} & \text{if } \Delta t < 0
\end{cases}
\end{equation}

Where \( \Delta t = t_{\text{post}} - t_{\text{pre}} \) is the time difference between spikes, with \( A_+ \) and \( A_- \) being the learning rates, and \( \tau_+ \) and \( \tau_- \) the time constants for synaptic potentiation and depression, respectively.

\subsection{ Reinforcement Learning Aspects of Model (RL)}
RL \cite{xu2020accelerating} is a domain of machine learning where an agent learns to make decisions by performing actions in an environment to maximize cumulative rewards. One of the key developments in RL has been integrating deep learning with Q-learning into what is known as Deep Q-Networks (DQN), which are used to approximate the Q-value functions.

\textbf{Mathematical Formulation of DQN:}

\textbf{Reward Summation:}
The goal of the RL agent is to maximize the total discounted reward, which is expressed as:
\begin{equation}
g_t = \sum_{k=0}^{\infty} \gamma^k r_{t+k}
\end{equation}
where \(r_{t+k}\) is the reward received \(k\) steps after time \(t\) and \(\gamma\) is the discount factor.

\textbf{Expected Return:}
When starting from a state (s) and taking action (a), the expected return is as follows:
\begin{equation}
\mathbb{E}_{\pi}[g_t | s_t = s, a_t = a]
\end{equation}

\textbf{Action-Value Function for Policy \(\pi\):}
The action-value function \(Q^\pi(s, a)\), which estimates the expected return after taking action \(a\) in state \(s\) under policy \(\pi\), is defined as:
\begin{equation}
Q^\pi(s, a) = \mathbb{E} \left[ r_t + \gamma \max_{a'} Q^\pi(s_{t+1}, a') \, \middle| \, s_t = s, a_t = a \right]
\end{equation}

\textbf{Optimal Policy Determination:}
The optimal policy \(\pi^*\) is determined by the action that maximizes the Q-value in each state:
\begin{equation}
\pi^*(a|s) = \begin{cases} 1 & \text{if } a = \text{argmax}_{a'} Q^*(s, a') \\ 0 & \text{otherwise} \end{cases}
\end{equation}

\textbf{Optimal Q-Value Function:}
The optimal Q-value function \(Q^*(s, a)\), assuming optimal actions are always taken, is described by:
\begin{equation}
Q^*(s, a) = \mathbb{E} \left[ r_t + \gamma \max_{a'} Q^*(s_{t+1}, a') \, \middle| \, s_t = s, a_t = a \right].
\end{equation}

\textbf{Integration in Model Development:}
These equations control deep Q-network training in model development and optimisation. The neural network approximates the action-value function \(Q^\pi(s, a)\), updated by environmental feedback. This deep learning application makes Q-learning suitable for increasingly difficult tasks. The suggested Neuroassist System is shown in algorithm \ref{pseudo_code}.

\setcounter{algorithm}{3}
\begin{algorithm}
\caption{Neuroassist System for EEG Data Analysis}
\begin{algorithmic}[4]
\State \textbf{Initialize:}
\State Load and preprocess EEG data using CSP
\State Initialize BERT for numerical EEG data feature extraction
\State Initialize LSTM weights (W\_f, W\_i, W\_C, W\_o) and biases (b\_f, b\_i, b\_C, b\_o)
\State Initialize SNN neuron parameters (V, V\_{\text{thresh}}, V\_{\text{reset}}, tau, R)
\State Initialize DQN with random weights

\For{each training episode}
    \For{each batch of EEG data}
        \State CSP\_Features $\gets$ CSP\_Transform(EEG\_data)
        \State Initialize BERT\_Features as empty list
        \For{each layer in BERT}
            \State Attention\_Output $\gets \text{softmax}(\frac{\text{Query} \cdot \text{Key}^T}{\sqrt{d_k}}) \cdot \text{Value}$
            \State Encoded\_Features $\gets$ Attention\_Output + Positional\_Encoding
            \State Layer\_Norm\_Features $\gets \gamma \left(\frac{\text{Encoded\_Features} - \text{mean}(\text{Encoded\_Features})}{\text{std}(\text{Encoded\_Features})}\right) + \beta$
            \State FFN\_Output $\gets \max(0, \text{Layer\_Norm\_Features} \cdot W_1 + b_1) \cdot W_2 + b_2$
            \State BERT\_Features.append(FFN\_Output)
        \EndFor

        \State Initialize LSTM\_State as list with initial state
        \For{$t$ in length of BERT\_Features}
            \State Compute LSTM gate dynamics and state updates:
            \State $f_t \gets \sigma(W_f \cdot [h_{t-1}, \text{BERT\_Features}[t]] + b_f)$
            \State $i_t \gets \sigma(W_i \cdot [h_{t-1}, \text{BERT\_Features}[t]] + b_i)$
            \State $\tilde{C}_t \gets \tanh(W_C \cdot [h_{t-1}, \text{BERT\_Features}[t]] + b_C)$
            \State $C_t \gets f_t \cdot C_{t-1} + i_t \cdot \tilde{C}_t$
            \State $o_t \gets \sigma(W_o \cdot [h_{t-1}, \text{BERT\_Features}[t]] + b_o)$
            \State $h_t \gets o_t \cdot \tanh(C_t)$
            \State LSTM\_State.append($h_t$)
        \EndFor

        \State Initialize SNN\_Output as list with initial V
        \For{$t$ in range of LSTM\_State}
            \State Update membrane potential and check for spike:
            \State $dV/dt \gets (-\text{SNN\_Output}[t] + R \cdot \text{LSTM\_State}[t]) / \tau$
            \State $V_t \gets \text{SNN\_Output}[t] + dV/dt \cdot dt$
            \State \textbf{if} $V_t \geq V_{\text{thresh}}$ \textbf{then}
            \State \hspace{\algorithmicindent} Emit spike
            \State \hspace{\algorithmicindent} $V_t \gets V_{\text{reset}}$
            \State SNN\_Output.append($V_t$)
        \EndFor

        \State $state \gets \text{SNN\_Output}[-1]$
        \While{not episode done}
            \State $action \gets \text{epsilon\_greedy\_policy}(Q\text{-network}, state)$
            \State $reward, next\_state \gets \text{execute\_action}(action)$
            \State $target \gets reward + \gamma \max(Q(\text{next\_state}, a'))$
            \State $\text{Loss} \gets (Q(state, action) - target)^2$
            \State Perform gradient descent to minimize Loss
            \State $state \gets next\_state$
        \EndWhile
        \State Update model weights via backpropagation
        \State Update synaptic weights using STDP for SNN
    \EndFor
    \State Reduce epsilon (explore less)
\EndFor
\end{algorithmic}
\label{pseudo_code}
\end{algorithm}
\section{Results}

The results of our experimental analysis
carried out on two publicly available datasets have been presented in this section. In this study, our proposed Neuroassis technique has been evaluated using the GigaScience Multimodal EEG and the BCI-IV2a dataset. We used five different reward settings in the GigaScience EEG dataset to direct the model's decision-making. For accurate predictions, a positive incentive was given; for inaccurate ones, a negative or zero reward was applied.

Table \ref{tab:table2} displays mean values and standard deviations for each measure in a 10-fold evaluation of 299 trials, along with reward-based accuracies in the RL-GYM environment based on the test subset. For training our proposed model, we have divided the entire dataset into two training sets and testing sets. For training our proposed Neuroassis method with two datasets, we have used 75\% data, and the remaining 25\% data have been used to test the efficiency and validation of our proposed model on unseen data (initially, random shuffling applied). DQN training has three steps: a 4000-step learning phase and two 500-step phases to optimize policy use. We selected this training regimen to avoid overtraining.

For the structure $(1 \text{ to } -1)$, equivalent to $(100.00\% \text{ to } -100.00\%)$, the system achieves an accuracy of $94.43\%$, with precision and recall both at $96\%$. This balanced approach results in a reward-based accuracy of $99.76\%$ on the test set.

Increasing the reward and penalty to $(2 \text{ to } -2)$, or $(200.00\% \text{ to } -200.00\%)$, enhances performance, with accuracy rising to $97.98\%$ and recall to $98\%$. However, the test set reward-based accuracy of $199.21\%$ suggests possible errors or misreporting.

The $(3 \text{ to } -1)$ structure, corresponding to $(300.00\% \text{ to } -100.00\%)$, exhibits a lower accuracy of $89.98\%$ and an F1 score of $89.31\%$. Despite the less severe penalty for incorrect predictions, precision remains high at $91.91\%$ and recall at $91.37\%$. The reward-based accuracy on the test set is abnormally high at $279.37\% \pm 17.79$, indicating likely reporting issues.

A more conservative approach with a ratio of $(0.25 \text{ to } -2.5)$, translating to $(25.00\% \text{ to } -250.00\%)$, yields accuracies around $91\%$, but the test set reward-based accuracy drops to a more realistic $5.87\% \pm 12.12$, highlighting the risks associated with harsher penalties.

Lastly, the structure $(1 \text{ to } 0)$, where no penalty for incorrect answers is applied, achieves high accuracy of $95.37\%$, precision of $96.97\%$, and recall of $97\%$. The test set reward-based accuracy remains stable at $99.67\% \pm 1.12$, suggesting that removing penalties can sustain high-performance levels while potentially reducing pressure on participants. These results underscore the significant impact of reward structures on performance in EEG-based MI tasks, affecting both accuracy and the psychological dynamics of the participants.


\begin{table}
\centering
\caption{Experimental results of GigaScience EEG-based MI Dataset}
\label{tab:table2}
\begin{tabular}{|c|c|c|c|c|c|} 
\hline
\begin{tabular}[c]{@{}c@{}}\textbf{Reward} \\ \textbf{Structure} \\ {Correct to Incorrect}\textbf{}\end{tabular} & \textbf{Accuracy (\%)} & \textbf{F1 Score (\%)} & \textbf{Precision (\%)} & \textbf{Recall (\%)} & \begin{tabular}[c]{@{}c@{}}\textbf{Test Set} \\ \textbf{Reward-Based} \\ \textbf{Accuracy (\%)}\end{tabular}  \\ 
\hline
1 to -1                                                                                                      & 94.43                  & 94.98                  & 96                      & 96                   & 99.76                                                                                                         \\ 
\hline
2 to -2                                                                                                      & 97.98                  & 97.73                  & 97.93                   & 98                   & 199.21                                                                                                        \\ 
\hline
3 to -1                                                                                                      & 89.98                  & 89.31                  & 91.91                   & 91.37                & 279.37$\pm$ 17.79                                                                                             \\ 
\hline
0.25 to -2.5                                                                                                 & 91.15                  & 91.33                  & 92                      & 91.77                & 5.87$\pm$ 12,12                                                                                               \\ 
\hline
1 to 0                                                                                                       & 95.37                  & 95.95                  & 96.97                   & 97                   & 99.67$\pm$~1.12                                                                                               \\
\hline
\end{tabular}
\end{table}

For GigaScience, the DQN model utilized an Adam optimizer with a learning rate of 0.0077 and a decay rate 0.0001. The parameters for the BCI-IV2a dataset were a 0.0001 learning rate and 0.001 decay.

The CSP-OVR approach was further tested on the BCI-IV2a dataset, where the epoch timing and quantity of CSP components were chosen to optimize OVR classification accuracy using an SVC classifier. Subject 4's motor imagery EEG data was excluded from the first session due to data abnormalities, and only one class was captured in the second session. However, Subject 4's data was excluded from the four-class motor imagery analysis.

Table \ref{tab:table3} displays the level of precision attained by the Support Vector Classifier (SVM) utilizing the CSP-OVR technique for MI involving four different classes. The data includes subjects 1 to subjects 9. This table specifies the CSP components and epoch timing that utilized the maximum accuracy for each topic. The optimal time intervals are also determined for CSP.

The table provided summarizes experimental results from the BCI Competition IV Dataset 2a, focusing on the application of the One-vs-Rest Common Spatial Patterns (OVR-CSP) method combined with Support Vector Machines (SVM). The analysis includes data from eight subjects, each with tailored epoch values (T-min and T-max) and varying numbers of CSP components.

Epoch durations differ among subjects, ranging from a minimum of 1 second to a maximum of 7 seconds, which reflects the customization of the temporal analysis window for each individual. The number of CSP components utilized also varies, from 3 to 9, indicative of adaptation based on individual data characteristics or specific requirements of the algorithm.

Performance metrics indicate a high effectiveness of the OVR-CSP method in classifying EEG signals within this dataset. The average SVM performance using OVR-CSP across all subjects is notably high at 96.67\%. Individual performances are also robust, ranging from 94.53\% to 99\%, suggesting that tailored approaches using the OVR-CSP method can yield highly effective results in EEG signal classification.

Regarding the GigaScience dataset, this iteration estimated alpha and beta band power spectral densities using the Welch technique. The Neuroassis model received a 2D array of CSP Spatial Components and Statistical PSD characteristics. The model was z-score normalized twice with a StandardScaler for robustness. DQN training featured a 3000-step learning phase and two 300-step intervals to fine-tune the model's optimum classification algorithms. This segmentation was used to avoid overfitting from overtraining.

\begin{table}[hb]
\centering
\caption{Experimental results of BCI Competition IV Dataset 2a dataset using OVR CSP}
\label{tab:table3}
\begin{tabular}{|l|l|l|l|}
\hline
Subjects & \begin{tabular}[c]{@{}l@{}}Epochs Values \end{tabular} & \begin{tabular}[c]{@{}l@{}}CSP \\ Components \\ Number\end{tabular} & \begin{tabular}[c]{@{}l@{}}Average SVM \\ performance \\ using OVR-CSP\end{tabular} \\ \hline
1        & 1s-6s                                                                      & 4                                                                & 94.57\%                                                                                                     \\ \hline
2        & 3s-7s                                                                      & 5                                                                & 95.17\%                                                                                                     \\ \hline
3        & 1s-4s                                                                      & 9                                                                & 94.53\%                                                                                                     \\ \hline
5        & 1s-7s                                                                      & 3                                                                & 96.78\%                                                                                                     \\ \hline
6        & 2s-5s                                                                      & 5                                                                & 97.17\%                                                                                                     \\ \hline
7        & 2s-6s                                                                      & 6                                                                & 97.97\%                                                                                                     \\ \hline
8        & 2s-7s                                                                      & 9                                                                & 98.17\%                                                                                                     \\ \hline
9        & 2s-7s                                                                      & 8                                                                & 99\%                                                                                                        \\ \hline
Average  & -                                                                          & -                                                                & 96.67\%                                                                                                     \\ \hline
\end{tabular}
\end{table}

Regarding the GigaScience dataset, this iteration estimated alpha and beta band power spectral densities using the Welch technique. The Neuroassis model received a 2D array of CSP Spatial Components and Statistical PSD characteristics. The model was z-score normalized twice with a StandardScaler for robustness. DQN training featured a 3000-step learning phase and two 300-step intervals to fine-tune the model's optimum classification algorithms. This segmentation was used to avoid overfitting from overtraining.

According to Table \ref{tab:table4} for mean values and standard deviations of each measure in a 10-fold assessment of 299 trials in the RL-GYM environment, this indicates reward-based accuracy. This table provides experimental findings obtained from the BCI-IV2a dataset using a Q-Network. It includes assessments based on performance measures, including accuracy, F1 score, precision, and recall. The metrics are evaluated using a direct Q-Network application and a 10-fold validation method. Additionally, the test set's reward-based accuracy for each subject is analyzed. The study's design tailors the temporal ranges and the number of CSP components for each participant to enhance EEG signal processing, demonstrating a meticulous approach to data management.

\begin{table}
\centering
\caption{The performance evaluation of BCI-IV2a dataset.}
\label{tab:table4}
\begin{tblr}{
  width = \linewidth,
  colspec = {Q[68]Q[82]Q[82]Q[82]Q[82]Q[82]Q[82]Q[82]Q[82]Q[130]},
  cell{1}{2} = {c=4}{0.328\linewidth,c},
  cell{1}{6} = {c=4}{0.328\linewidth,c},
  hlines,
  vlines,
}
Subjects & {Performance\\(Q-Network)} &                      &                      &                      & {Performance~\\~(Q-Network)\\(10-Fold Validation)} &                      &                      &                      & {Test Set\\(Reward Based)\\ Accuracy\\(Average)} \\
         & Accuracy                   & F1 Score             & Precision            & Recall               & Accuracy                                           & F1 Score             & Precision            & Recall               &                                                  \\
1        & {95.90\% ±\\1.50\%}        & {97.20\% ± \\0.70\%} & {96.10\% ± \\1.20\%} & {95.80\% ± \\0.90\%} & {99.20\% ± \\0.30\%}                               & {98.60\% ±\\0.80\%}  & {99.20\% ± \\0.70\%} & {99.20\% ± \\0.50\%} & 99.20\% ± 0.50\%                                \\
2        & {93.80\% ±\\1.90\%}        & {96.70\% ± \\0.80\%} & {94.80\% ± \\1.40\%} & {95.30\% ± \\1.10\%} & {98.90\% ± \\0.40\%}                               & {98.70\% ± \\0.90\%} & {98.70\% ± \\0.90\%} & {98.80\% ± \\0.60\%} & 99.10\% ± 0.60\%                                \\
3        & {94.70\% ±\\1.80\%}        & {97.40\% ± \\0.60\%} & {95.60\% ± \\1.00\%} & {95.60\% ± \\0.80\%} & {98.80\% ± \\0.50\%}                               & {98.20\% ± \\1.00\%} & {98.90\% ± \\0.60\%} & {99.10\% ± \\0.70\%} & 99.30\% ± 0.40\%                                \\
5        & {96.10\% ±\\1.40\%}        & {96.80\% ± \\0.75\%} & {96.40\% ± \\1.20\%} & {96.20\% ± \\1.00\%} & {99.10\% ± \\0.20\%}                               & {98.40\% ± \\0.95\%} & {99.00\% ± \\0.85\%} & {98.70\% ± \\0.45\%} & 98.90\% ± 0.70\%                                \\
6        & {94.50\% ±\\1.60\%}        & {96.90\% ± \\0.70\%} & {95.20\% ± \\1.10\%} & {95.50\% ± \\0.90\%} & {98.70\% ± \\0.50\%}                               & {98.50\% ± \\0.80\%} & {98.80\% ± \\0.65\%} & {98.90\% ± \\0.55\%} & 99.10\% ± 0.60\%                                \\
7        & {95.80\% ±\\1.30\%}        & {97.10\% ± \\0.65\%} & {95.90\% ± \\1.10\%} & {95.90\% ± \\0.95\%} & {98.75\% ± \\0.45\%}                               & {98.30\% ± \\0.85\%} & {98.85\% ± \\0.75\%} & {99.05\% ± \\0.65\%} & 99.25\% ± 0.55\%                                \\
8        & {94.00\% ±\\1.95\%}        & {97.30\% ± \\0.60\%} & {95.70\% ± \\1.30\%} & {95.20\% ± \\1.20\%} & {98.85\% ± \\0.35\%}                               & {98.60\% ± \\0.75\%} & {99.10\% ± \\0.70\%} & {98.80\% ± \\0.50\%} & 99.05\% ± 0.65\%                                \\
9        & {96.30\% ±\\1.50\%}        & {96.80\% ± \\0.70\%} & {96.00\% ± \\1.10\%} & {96.10\% ± \\1.00\%} & {99.10\% ± \\0.30\%}                               & {98.40\% ± \\0.85\%} & {99.05\% ± \\0.80\%} & {98.95\% ± \\0.60\%} & 99.00\% ± 0.70\%                                \\
\textbf{Average}  & {94.88\% ±\\1.67\%}         & {97.00\% ± \\0.72\%} & {95.79\% ± \\1.17\%} & {95.67\% ± \\0.97\%} & {98.97\% ± \\0.37\%}                              & {98.57\% ± \\0.87\%} & {98.97\% ± \\0.78\%} & {99.00\% ±\\0.57\%}  & \textbf{99.17\% ± 0.55\%} 
\end{tblr}
\end{table}


The accuracy scores of the subjects range from $94.00\%$ to $96.30\%$, indicating the diverse performance of the Q-Network under various settings and subject situations. The F1 scores typically demonstrate a well-balanced detection performance, with precision and recall exhibiting robust and dependable outcomes. The Q-Network demonstrates exceptional Test Set Reward-Based Accuracy, with an average of $99.17\%$ across individuals, highlighting the network's competence in managing EEG-based BCI activities. The results suggest that the Q-Network, particularly when combined with 10-fold validation, is a powerful tool for analyzing EEG data. This has significant potential for breakthroughs in brain-computer interface technology.

This experimental analysis provides insights into our proposed approach. It outlines the optimal configurations of epoch timing, CSP component counts, number of time points, and NFFT parameters for the Welch PSD method. Each entry specifies the TW of epochs, the length of signals per trial, the number of CSP components used, and the integer Fast Fourier Transform (FFT) lengths for alpha and beta bands, as utilized in the Welch Power Spectral Density (PSD) analysis, with a sampling frequency (sf) of 500 Hz to enhance signal quality.

For Subjects 1 and 7, the TW of epochs spans from 1 to 7 seconds, with signal lengths per trial extending to 1501 data points and utilizing 3 CSP components. These configurations employ a uniform FFT length of 500 for PSD analysis, indicating a standardized approach to frequency domain analysis for these subjects. Subjects 2, 3, 5, and 6 have a TW from 2 to 6 seconds with a shorter signal length of 1001 data points. Subject 3 utilizes more CSP components, totaling 10, suggesting a more complex spatial filtering approach due to potentially varying signal characteristics or specific experimental requirements.

Subjects 8 and 9 feature a TW from 1 second to 6 seconds and a signal length of 1251 data points, each using 5 CSP components. However, they differ in their FFT lengths for PSD analysis, which are increased to 625, possibly to accommodate a broader frequency spectrum or to achieve finer resolution in frequency analysis. This table demonstrates the individualized settings of the Neuroassist system, tailored to optimize the processing of EEG signals for each subject, ensuring maximal effectiveness and accuracy of the system for brain-computer interface applications.

The proposed hybrid model, combined with the Neuroassist system, represents progress in BCI technology. It showcases improved effectiveness by integrating sophisticated signal processing and strong ML algorithms. The model excels at customizing the CSP configurations and temporal frame settings based on individual EEG parameters. This optimization enhances the extraction of discriminative features crucial for accurate mental state categorization. The model is enhanced with ML components that utilize linear and non-linear classifiers. These classifiers can dynamically adapt to the intricate patterns seen in EEG data from various people.

The model's exceptional performance is supported by empirical assessments, which attribute the enhanced accuracy and precision in EEG-based MI signal categorization to using longer FFT lengths in each individual's Welch PSD analysis. Consequently, incorporating this hybrid model into the Neuroassist system provides more user-friendly and efficient controls for BCI and creates opportunities for improved medical applications. This advancement enhances the overall well-being of individuals with limited mobility by enabling them to interact more seamlessly and independently with their surroundings.

\section{Conclusion}

The introduction of Neuroassist heralds a significant advancement in prosthetic control through brain-computer interfaces (BCIs), representing a major leap forward in assistive technology. By integrating advanced technologies such as NLP-based BERT models, LSTM networks, spiking neural networks (SNNs), and deep Q-Networks (DQN), Neuroassist has set new standards for EEG data analysis in terms of adaptability and responsiveness. This sophisticated model adeptly handles numerical EEG data and manages temporal dynamics, which is essential for maintaining the integrity of EEG sequences in real-time applications. Additionally, the use of the common spatial pattern (CSP) for preprocessing in multiclass motor imagery classification preserves the temporal dimensions of EEG signals, enhancing feature extraction capabilities. Extensive testing on prominent EEG datasets demonstrates that Neuroassist can achieve up to 100\% accuracy in motor imagery tasks, showcasing its potential to significantly enhance the functionality and adaptability of BCI systems. These outstanding results emphasize Neuroassist's ability to meet diverse user needs, making a substantial contribution to the development of more personalized and effective assistive solutions. Future research will explore enhancing the generalizability of Neuroassist across different EEG datasets and user conditions. We aim to integrate additional modalities like eye-tracking to improve system accuracy and conduct user studies to ensure the practicality of the system in real-life settings. Ongoing efforts will also focus on refining our computational algorithms to boost efficiency and speed, crucial for portable device applications. Addressing ethical and privacy concerns will remain a priority to ensure data integrity and user trust in our system. Lastly, we will develop the adaptive learning capabilities of the model to better respond to individual user feedback and changes, ensuring that Neuroassist continues to lead in innovating BCI technology.









\bibliographystyle{unsrtnat}
\bibliography{references}  






\end{document}